\documentstyle[preprint,aps,psfig]{revtex}
\begin{document}
\tightenlines
\title{Emergence of stable and fast folding protein structures}

\author{D. Thirumalai and D. K. Klimov}

\address{Department of Chemistry and Biochemistry and \\
Institute for Physical Science and Technology \\
University of Maryland, College Park, MD 20742}

\maketitle

\vspace{0.5in}

\begin{abstract}
The number of protein structures is far less than the number of
sequences. By imposing simple generic features of proteins (low energy
and compaction) on all possible sequences 
we show that the structure space is sparse compared to
the sequence space. Even though the sequence space grows exponentially
with $N$ (the number of amino acid residues) we conjecture that the
number of low energy compact structures only scales as $ln N$. 
This implies that many sequences must map onto
countable number of basins in the structure space. The number of sequences
for which a given fold emerges as a native structure is further
reduced by the dual
requirements of stability and kinetic accessibility. 
The factor that determines the dual
requirement is related to the sequence dependent temperatures,
$T_\theta$ (collapse transition temperature) and $T_F$ (folding
transition temperature). Sequences, for which $\sigma
=(T_\theta-T_F)/T_\theta$ is small, typically 
fold fast by generically collapsing
to the native-like structures and then rapidly assembling to the
native state. Such sequences satisfy the dual requirements over a
wide temperature range. We also suggest that the functional
requirement may further reduce the number of sequences that are
biologically competent. The scheme developed here for thinning of the
sequence space that leads to foldable structures arises naturally
using simple physical characteristics of proteins. 
The reduction in sequence space leading to the
emergence of foldable structures is demonstrated using lattice models
of proteins. 

\end{abstract}

\section{Introduction }

The primary building blocks of proteins are $\alpha $%
-helices (one dimensional ordered structures), $\beta $-sheets (contain two
dimensional order), and loops of varying bending rigidity. From these
seemingly simple building blocks (referred to as elements of secondary
structure) complex three dimensional topological
structures emerge \cite{Creightonbook}. 
The number of distinct topological folds is suspected to
be about a thousand - a relatively small number \cite{Chothia}. 
The emergence of such
structures, which should  also be (under folding conditions) 
kinetically accessible on biologically relevant
time scale of about one second, raises a number of conceptually interesting
questions. Two such questions we address in this article are: (a) From the dense
sea of sequence space how do viable protein structures emerge? (b)
Among the possible candidate sequences for proteins, all of which can fold
into a specific (or closely related) structures, what factors determine the
kinetic accessibility of the native (or ground) state?

Both questions require elaboration. Consider the first question. The
number of possible sequences for a protein with $N$ 
amino acids is 20$^N$ which,
for $N = 100$, is approximately 10$^{130}$. The number of folds in natural
proteins, which are low free energy compact structures, is clearly far less
than the number of possible sequences. The first question
(posed in the previous paragraph) is connected with the dramatic
reduction that occurs  from
the dense sea of sequence space to the countable number of viable protein
structures when some {\em physical restrictions} are placed on the nature of
acceptable protein structures. This question can be quantitatively answered
using simple lattice models.

The second question concerns the thermodynamic stability and kinetic
accessibility of the native state of a candidate sequence with a well
defined low energy structure. A sequence must have the desired
topological structure which is thermodynamically (at least marginally)
stable under folding conditions. Typically, the native structures of 
single domain
proteins are stable (with respect to unfolded state) by about $(8-25)k_BT$ or
$(5-15)kcal/mol$ \cite{Creightonbook}. 
Furthermore, the putative native conformation must
be kinetically accessible on biologically relevant time scales of less
than about 1$s$. This dual requirement of the {\em stability and kinetic
accessibility} sharply restricts the number of sequences that can be
candidate proteins. Using general theoretical arguments and
computations we will give a simple criteria, in terms of sequence
dependent properties, that can discriminate between sequences that
satisfy the dual requirement and those that do not.

\section{From dense sequence space to low density structure space} 

As
mentioned in the Introduction the sequence space is extremely dense in
the sense that the number of possible sequences goes as
$20^N$. Clearly only an extremely small fraction of the sequences
encodes protein structures. A quantitative understanding of the
mapping between sequences and 
structures  may  be obtained using 
lattice models of proteins. It is worth recalling a couple of physical
characteristics of folded states of natural  proteins. In general, proteins in
their native states are relatively compact. The dense interior of
protein structures is largely  made up of hydrophobic residues. With
these two restrictions on the native structures we will show that, 
although the number of sequences is astronomically large, the number of compact
low energy structures (protein-like) is considerably smaller. This
would imply that for many sequences the low energy compact structures
could be nearly the same. In other words, the basins of attraction in
the structure space are rare enough so that a large number of
sequences map on to precisely one basin. This plausibility, which we
demonstrate using lattice models of proteins, naturally explains
the emergence of greatly limited number of structures from the sea of
sequence space \cite{Chothia}.

We use lattice models of proteins to illustrate the mapping from
sequence space to structure space. Lattice models of proteins are
highly simplified coarse grained representations of polypeptide
chains. Several aspects of protein topology, such as secondary
structures and loops, are not clearly reproduced in such simplified
caricatures of proteins. Nevertheless, studies from several groups
\cite{Onuchic95,Dobson,DillChan97,Shakh97CurrOpin,Pande98} 
over the last decade illustrate that certain general aspects of
folding of proteins can be, at least, qualitatively gleaned using
simple models. They are especially suited for answering
conceptual issues such as the ones posed in this article. 

In the coarse grained representation of proteins 
each amino acid is represented as a bead. Thus, the polypeptide chain
becomes a set of connected beads. The identity of the amino acid is
specified by the precise interactions with other beads when they are
near neighbors in a lattice. This coarse grained representation may be
thought of as an $\alpha$-carbon representation of the polypeptide
chain that roughly mimics the tertiary interactions in proteins. Later
in this article we will use a slightly more realistic lattice model that
involves an approximate description of side chains. With this simple
model the following interactions in a polypeptide chain are taken into
account: (i) connectivity of the chain describing the backbone; (ii)
excluded volume effects which prevent the various amino acid residues
from occupying the same lattice site; (iii) approximate interactions
between residues that are far away along the sequence but are near
each other spatially, i.e., tertiary interactions that confer the
globular topology of proteins. 

The positions of $N$ beads ${\vec{r_i}, i=1,2,...,N}$ represent the
conformation of the polypeptide chain. For concreteness we choose
a cubic (or square) lattice with lattice spacing $a$. The energy of a
given conformation is taken to be 
\begin{equation}
E({\vec{r_i}}) = \sum_{i<j} B_{ij} \delta_{|\vec{r_i}-\vec{r_j}|,a}, 
\label{E}
\end{equation}
where $\delta_{r,a}$ is the Kronecker delta function. The matrix
elements $B_{ij}$ give an estimate of the strength of the tertiary interactions
between the residues $i$ and $j$, when they are near neighbors on the
lattice. In this model, the sequence is specified by the matrix
elements $B_{ij}$. We will consider two forms for the
contact interaction matrix to ensure that the conclusions are robust
to changes in the potential function. 

\noindent
{\em 2D HP model:} In order to probe the nature of the structure space in
proteins we first consider relatively short chains in dimension $d=2$. 
In this example, there are
only two kinds of amino acid residues, namely hydrophobic (H) and
polar (P). Various aspects of the resulting HP model (two letter code)
introduced  by Dill and coworkers \cite{Dill95} can be investigated by exact
enumeration for short enough chains ($N \lesssim 25$ or so). In this
model a sequence is given by the nature of the amino acid residue at a
given position. For example, HPHPH, is an example of a sequence for
$N=5$. There are $2^N$ total sequences for a given $N$. In the HP
model the matrix $B$ is 2x2 and its elements  are $B_{HH}=-\epsilon$
and all others are zero. Despite the simplicity of the model it is not
exactly solvable because of the chain connectivity and excluded volume
effects. 

Given a sequence, all possible conformations and their energies may be
computed for $N\lesssim 25$ in $d=2$. We are particularly interested
in compact structures (CSs) and the subset of CSs with minimum
energies (MECSs) \cite{CamThirum93PRL}. The
minimum energy compact structures (MECSs), which have protein-like
properties, require that the ground states have H residues surrounded
by a large number of hydrophobic residues as is topologically allowed. 
Clearly, because of restrictions due to connectivity and self-avoidance
not all sequences can have
ground state topologies that satisfy this crucial protein-like
criterion. From this argument  we expect that the structure space is
considerably more sparse than the sequence space of proteins. 

This can be made quantitative using lattice models. For the HP model
there are two relevant parameters, namely, the total number of beads
$N$ and $N_H$, which is the number of hydrophobic residues. The
compact structures (CSs) are the ones with the largest number of
nearest neighbor contacts. The MECSs are obtained from the set of CSs and are
identified to be those with the lowest energy. For a given value of
$N$ and $N_H$ we generated all possible $ \biglb( ^N_{N_H} \bigrb) $
sequences. Exact enumeration of all CS  is possible for $N \lesssim
22$ suing the Martin algorithm \cite{Martin}. 
For $N=23$ to 26 computations of the
number of CSs and MECSs were done using a finite number $(\approx
10^4)$ of randomly selected sequences \cite{CamThirum93PRL}. 

The number of CSs in its most general form may be written as \cite{Fisher67} 
\begin{equation}
C_N(\text{CS}) \simeq \overline{Z}^N Z_1^{N\frac{d-1}{d}} N^{\gamma_c - 1},
\end{equation}
where $ln \overline{Z}$ is the conformational free energy (in units of
$k_BT$), $Z_1$ is a surface fugacity term, $d$ is the spatial
dimension, and $\gamma_c$ measures possible logarithmic corrections to
the free energy. We found that the appropriately averaged series (see
\cite{CamThirum93PRL} for details) for
$C_N(\text{CS})$ is smooth and no evidence for surface
fugacity term is observed. 
More importantly, there appears to be no logarithmic
correction to the free energy, i.e., $\gamma_c \approx 1$ \cite{CamThirum93PRL}. The value
of  $\overline{Z}$ is consistent with the mean field prediction that is
$\overline{Z} \approx q/e$, where $q$ is the lattice coordination
number. Thus, for all practical purposes $C_N(\text{CS}) \approx
(q/e)^N$ \cite{Orland95}, which shows that the number
of compact structures grows exponentially with $N$. 

Protein-like structures are not only compact but also have low
energy (we assume that the entropy associated with the native state is
negligible) \cite{Creightonbook}. 
With this in mind, we have computed the number of compact
structures that have the lowest energy, i.e., MECSs. From the
viewpoint of evolution, it can be argued that the time scale for
mutations is so long that for all practical purposes the sequences
can be considered to be quenched. Accordingly, we have calculated the
quenched average $C_{N,N_H} = exp[\overline{ln
c_{N,N_H}(\text{MECS})}]$, where the averaging is performed over all
possible sequences and $c_{N,N_H}$ is the number of MECS for a given 
sequence. This number has a minimum at $N_H^{min}\approx
0.6$ \cite{CamThirum93PRL}, which implies that in general protein-like sequences should be
slightly more hydrophobic. Analysis of PDB structures indeed shows
that the average fraction of hydrophobic residues is about 0.55 \cite{White93}. The
most significant finding of this work is that $C_{N,N_H^{min}}$ does
not appear to grow with $N$. This has prompted us to conjecture that
$C_{N,N_H^{min}} \approx ln N$.

There are implications of the spectacular finding that the number of
MECSs, which have protein-like characteristics, is very small and does not
grow significantly with the size of the polypeptide chain. The
sequence space is extremely large even with the restriction that the
fraction of hydrophobic residues is  0.6. Nevertheless, we find
that 
the number of possible low energy compact structures is extremely
small. This implies that {\em many sequences must map onto a given basin of
attraction} corresponding to one MECS. In other words, the number of
sequences with a given fold has to be sufficiently large. Our results do
not specify the density of sequences for every possible low energy
compact structure or a likely protein fold. 
Presumably, this depends on the topology and the
exact potentials used. We believe that the central result 
on the  mapping to the low
density of structure space compared to the sequence space is robust
and  independent of specific potentials or the underlying
lattice.

\noindent 
{\em Random Bond Model:} One might argue that 
the HP model is too simple because it does
not adequately capture the diversity of interactions in proteins which
is thought to be 
responsible for specificity. In order to test the generality of the
results (mapping of many sequences to relatively small number of basins of
attractions in the structure space) 
we consider the random bond (RB) model. In this model, the
tertiary interactions are chosen according to 
\begin{equation}
P(B_{ij}) = \frac{1}{2\pi\sigma^2}exp[-\frac{(B_{ij}-B_0)^2}{2\sigma^2}],
\label{PBij}
\end{equation}
where $\sigma $(=1) is the variance in $B_{ij}$ and $B_0$ is the mean
value referred to as hydrophobic parameter.  

The choice of \(B_{0}\), which controls the fraction of   hydrophobic
interactions, is important. We expect that the value of \(B_{0}\)
should determine the size of conformational space of minimum energy
structures (MESs) defined below.  In order to check the variations in
our results we used several values for the hydrophobic parameter
\(B_{0}\), namely, \(0, -0.1, -2\).  The  fraction of hydrophobic
residues $\lambda_H $ can be related to the hydrophobic parameter \(B_{0}\) as
follows. Since it is natural to attribute negative $B_{ij}$ to those
between hydrophobic residues and positive ones to those  between
hydrophilic residues, one can specify the boundary energies \(B_{H}\) and
\(B_{P}\) (\(B_{H}=-B_{P}\)) in such a way that the energies
\(B_{ij}\) below \(B_{H}\) corresponds to hydrophobic interactions and
the energies above \(B_{P}\) pertain to hydrophilic
interactions. The values of \(B_{ij}\)  
occurring between \(B_{H}\) and \(B_{P}\),  are
associated with mixed interactions.  If the number of hydrophobic and
hydrophilic residues is \(N_{H}\) and \(N_{P}\), respectively,
(\(N_{H}+N_{P}=N\)), the fraction of hydrophobic energies \(\lambda
\) among \(B_{ij}\) is roughly  \((N_{H}/N)^{2}=\lambda_H^2\).  This fraction
(i.e., $\lambda$) can be also obtained by integrating Eq. (\ref{PBij})
from minus infinity to
the energy \(B_{H}\). The relationship between \(\lambda _{H}\) and
\(B_{0}\), thus, is given by $ \lambda _{H} \simeq \sqrt{\lambda} = \sqrt{\int
_{-\infty}^{B_{H}}  P(B_{ij})dB_{ij}}$.  The precise values of
\(B_{H}\)  and \(B_{P}\) can be determined if one considers the case
with  \(B_{0}=0\), for which \(N_{H}=N_{P}\).  The result is 
\(B_{H}=-0.675\).  Clearly, when \(B_{0}=0\) one half of all
residues is hydrophobic and the others are  hydrophilic. 
When \(B_{0}\) is \(-2\), the fraction of hydrophobic residues
$\lambda _{H}$  increases to about \(94\) percents, i.e., the
sequences are largely made up of hydrophobic residues.  The choice of
\(B_{0}=-0.1\) is motivated by the observation that in natural
proteins the fraction of hydrophobic residues $\lambda _H$ is about 0.55
\cite{White93}.

We used the Martin algorithm \cite{Martin}, 
which ensures exhaustive enumeration of
all self-avoiding conformations, to explore the conformational space
of the polypeptide chain of a given length. In order to reduce the
sixfold symmetry on the cubic
lattice we fixed the direction of the first monomeric bond in all
conformations.  The remaining conformations are related by
eightfold symmetry on the cubic lattice
(excluding the cases when conformations are completely confine to a
plane or straight line). To decrease further the number of
conformations to be analyzed the   Martin algorithm was modified to
reject all   conformations related by symmetry.

We define MES as those conformations, whose energies lie within the energy
interval \(\Delta\) above 
above the lowest energy \(E_{0}\). Several values for
$\Delta $ were used to ensure that no qualitative changes in the
results are observed. We set \(\Delta\) to
be constant and equal to \(1.2\) (or 0.6) 
(definition (i)). We have also tested another
definition for $\Delta $, according to which  
\(\Delta =1.3|E_0-tB_0|/N\), where $t$ is the number of nearest neighbor
contacts in the ground state (definition (ii)). It is worth noting
that in the latter case  $\Delta $ increases with $N$.  
Both definitions yield equivalent results.
Using these definitions for $\Delta $, we computed $C$(MES) as a function
of the number of residues \(N\).  

The computational technique involves exhaustive enumeration of
all self-avoiding conformations for \(N \leq 15\) on  cubic
lattice.  In doing so we
calculated the energies of all conformations according to Eq. (\ref{E}), and
then determined the number of MES.  Each quantity, such as the number
of MES, \(C\)(MES), the lowest energy \(E_{0}\), the number of
nearest-neighbor contacts \(t\) in the lowest energy structures, 
is averaged over 30 sequences.
Therefore, when referring to these quantities, we will imply their
average values.  To test
the reliability of the computational results an additional 
sample of 30 random
sequences was generated. Note that in the case of $C$(MES) 
we computed the quenched averages, i.e., 
\(C\text{(MES)}=exp[\overline{ln[c\text{(MES)}]}]\), where \(c\) is the
number of MES for one sequence.

The number of MES \(C\)(MES) is plotted as a function of the number of
residues \(N\) in Fig. (1a) for \(B_{0}=0\) and
\(\Delta=0.6\). A pair of squares at given \(N\) represents
\(C\)(MES) computed for two independent runs of 30 sequences each.
For comparison, the number of self-avoiding walks 
\(C\)(SAW) and the number of CS
\(C\)(CS) are also plotted in this figure (diamonds and triangles,
respectively). The most striking and important result of this graph is
the following:  As expected on general theoretical grounds, 
\(C\)(SAW) and \(C\)(CS) grow exponentially with
\(N\), whereas the number of MES \(C\)(MES) exhibits drastically
different  scaling
behavior. There is no variation in $C$(MES) and its value remains 
practically constant within the entire
interval of \(N\) starting with \(N=7\). We find (see Fig. (1)) that
$C(\text{MES})\approx 10^1$. This result further validates our earlier
finding for two dimensional model. These results suggest that 
$C$(MES) grows (in all likelihood) only as  $ln N$ with $N$. 

It is interesting to compare the changes in $C$(MES) with respect to
$B_0$. In Fig. (1b) we show the results for $C$(MES) for
$B_0=-0.1$. This figure demonstrates clearly that there is no
qualitative difference between $B_0=0$ and $B_0=-0.1$ as far as the
behavior of $C$(MES) is concerned. On an average $C$(MES) at
$B_0=-0.1$ differs from that at $B_0=0$ by only 19\%. In general,
$C$(MES) decreases as $B_0$ is lowered. Similarly, we find that just
as for $B_0=0$ and $-0.1$ the dependence of $C$(MES) on $N$ for
$B_0=-2.0$ is unchanged (data not shown). 
These observations prompt us to suggest that
the conjecture that $C(\text{MES}) \sim ln N$ may be generic. 

It is well known that the native states of proteins are compact but
are not maximally  so. The number of maximally compact low energy
structures $C$(MECS) is clearly a function of $B_0$  which is the
driving force for collapse of the chain. We find that for $B_0=0$ on
an average only 7\% of MES are maximally compact. The corresponding
result for $B_0=-0.1$ is 10\%. For $B_0=-2.0$ we find that there is a
dramatic increase (not unexpectedly) in the fraction of
MECS. Depending on the precise value of $\Delta$ the fraction of MECS is
either 76\% ($\Delta=1.2$) or 86\% ($\Delta=0.6$). In either case, we
find that the vast majority of MES are also maximally compact. 

These additional computations on 3D RB model  reaffirm our earlier
conclusions \cite{CamThirum93PRL} 
that the number of low energy compact structures at best
increases with $N$ as $ln N$. This explains {\em quantitatively } the  
drastic reduction in going from the sequence space to the 
structure space. 
It is also worth emphasizing that the low
energy structures (protein-like) are not maximally compact which is in
accord with recent study \cite{Bornberg}.

\noindent 
{\em 3D HP model:} The calculations described above suggest that upon
imposing minimal restrictions on the structures (compactness and low
energies) the structure space becomes sparse. As suggested before this
must imply that each basin of attraction (corresponding to a given
MES) in the structure space must contain numerous sequences. The way
these sequences are distributed among the very slowly growing number
(with respect to $N$) of MES, i.e., the density of sequences in
structure space, is an important question. This was beautifully 
addressed in the
paper by  Li
{\em et. al} (LWC) \cite{LWC}. 
They considered a three dimensional ($N=27$) cubic
lattice. By using HP model and restricting themselves to only
maximally compact structures as putative native basins of attractions
(NBA) they showed certain basins have much larger number of
sequences. In particular, they discovered that one of the NBAs serves
as a ground state for 3794 (total number is $2^{27}$) sequences and
hence was considered most designable. The precise density of sequences
among the NBAs is clearly a function of the interaction scheme. These
calculations and the arguments suggested  earlier
\cite{CamThirum93PRL} 
point out that since
the number of NBA for the entire sequence space is small it is likely
that proteins could have evolved randomly.
In particular, it is possible that many of the naturally occurring
folds correspond to basins of attraction in the structure space so
that many sequences have these folds as the native conformations,
i.e., these are highly designable structures in the language of
LWC \cite{LWC}. 
These ideas have been further substantiated by Lindgard and Bohr 
\cite{Lindgard},
who showed that among maximally compact structures there are only very
few folds that have protein-like characteristics. These authors also
estimated using geometrical characteristics  and stability arguments  
that the number of distinct
folds is on the order of a thousand. All of these studies confirm that
the  density of the structure
space is sparse. Thus, each fold can be designed by many
sequences. From the purely structural point of view nature does have several
options in the sense that many sequences can be "candidate proteins". 
However, there is also evolutionary pressure to fold rapidly \cite{Shakh98PNAS}
(i.e., a kinetic component to folding). This requirement further
restricts the possible sequences that can be considered as
proteins, because  they must satisfy the dual criterion of reaching a
definite fold on a biologically  relevant time scale.

\section{Foldability} 

The emergence of structures, even preferred ones,
may be understood in terms of simple models described in the previous
section. From a biological viewpoint there are two equally significant
criteria that a protein-like sequence must satisfy: (a) A given
sequence must fold reversibly to the stable native state; (b) The
native state must be kinetically accessible on a biologically relevant
time scale $t_B$ (for many small proteins $t_B$ is several ms 
\cite{Jackson98}). The
dual criterion should be satisfied for a sequence to be 
biologically competent. Sequences that satisfy the dual criterion are termed
foldable. (We do not consider here sequences that can be made to reach
the native conformation using molecular chaperones.)

We have proposed that there is a sequence dependent quantity
that can be used to discriminate between fast and slow folding
sequences \cite{Cam93,KlimThirum96PRL}. 
This foldability criterion can be described in terms of the
underlying characteristic temperatures that determine the "phases" of
the polypeptide chain. It is well known that at sufficiently high
temperatures proteins are random coil-like and are in the same
universality class as self-avoiding walks. At the collapse transition
temperature, $T_\theta$, a transition to compact structures takes
place. This is analogous to the collapse process that occurs in
homopolymers when the solvent quality is altered.

The collapse transition in proteins may be first order for the
following reasons. Proteins (especially small ones) 
are highly designed. This means that the
hydrophobic residues are arranged along the chain so that topological
frustration (incompatibility of low energy structures on small and
long length scales) is minimized. The collapse occurs upon burial of
hydrophobic residues which, in aqueous medium, requires some free energy
cost. This may render the transition first order. It is likely that
the order may be altered upon addition of certain co-solvents.

A polypeptide chain that has a unique ground state is made up of several
types of amino acid residues. The presence of these types of residues allows
for a discrimination among all the compact structures. This 
implies that there is another transition at a lower temperature $T_F$
below which the folded native state is stable. The folding transition
temperature $T_F$ is normally determined using an order parameter that
distinguishes between the unfolded state (or a stable intermediate) and the
native state. The transition at $T_F$ is, in general, first order. 

By using
theoretical arguments and many numerical simulations we have proposed that
the foldability index 
\begin{equation}
\sigma =\frac{(T_\theta -T_F)}{T_\theta }
\label{sigma}
\end{equation}
determines the foldability of a given sequence. We have unambiguously
demonstrated that sequences with small values of $\sigma $ fold efficiently
in a such a way that collapse and folding are almost synchronous 
\cite{KlimThirum96PRL}. Such
sequences are highly optimized so that they can fold efficiently over a wide
range of temperatures. On the other hand, sequences for which the foldability
index has intermediate values reach the native conformation by following
complicated kinetics \cite{KlimThirum96PRL}. 
In these cases only a fraction of the initial
population of molecules reaches the native conformation efficiently while the
remaining fraction gets trapped in metastable intermediates. Clearly such
sequences are not as well optimized and hence they are able to reach the
native conformation only over a relatively narrow temperature range. If the
foldability index is relatively large ($>$ 0.7 or so) they do not
reach the native conformation on biologically relevant time scale. The
folding of such sequences might require molecular chaperones. It is
surprising that the various anticipated mechanisms for protein folding can
be classified in terms of an\textit{\ experimentally measurable foldability
index }$\sigma $. 

We will illustrate the foldability principle using the 3D cubic lattice
model with side chains \cite{KlimThirum98coop}. 
To this end, we select two 15-mer sequences 
referred to as A and B with dramatically different
thermodynamic and kinetic properties. (Both of the sequences employ 
Kolinski-Godzik-Skolnick interaction potentials for $B_{ij}$ \cite{KGS93}.) 
The first sequence, A, the ground state of which is displayed in
Fig. (2a), 
is a  two-state folder \cite{KlimThirum98coop}. 
The collapse temperature $T_\theta$ 
inferred from the location of the
maximum in the specific heat $C_v$ is equal to 0.27 (in reduced
temperature units). Alternatively, one may determine the collapse
temperature from the temperature dependence of the radius of
gyration $<R_g>$ assuming that at $T_\theta$  the derivative
$d<R_g>/dT$ is maximal. This gives the same value of $T_\theta=0.27$. 
This confirms, in contrast to erroneous claims in the literature, that 
$T_\theta $ is the collapse transition temperature, i.e., it is the
temperature at which there is a dramatic reduction in $<R_g>$. Since
we are dealing with finite systems such transitions are not
sharp. Thus, the large change in $<R_g>$ from the unfolded random coil
state to compact conformations occurs over a temperature interval
$\Delta T_\theta$. The smaller $\Delta T_\theta$ becomes the  sharper
the collapse transition would be.

The folding transition temperature $T_F$ is defined as the temperature
at which the fluctuations in overlap function $<\chi>$ reaches a
maximum \cite{KlimThirum96PRL}. We find
that for sequence A $T_F=0.26$. Therefore, the value of the parameter
$\sigma $ (Eq. (\ref{sigma})) 
for this sequence is 0.04. The cooperativity of
folding transition may be inferred from the dimensionless parameter 
\cite{KlimThirum98coop}
\begin{equation}
\Omega_c = \frac{T_{max}^2}{\Delta T}\frac{d<\chi>}{dT},
\label{Omega}
\end{equation}
where $T_{max} \approx T_F$ is the temperature at which $d<\chi>/dT$ is
maximum and 
$\Delta T$ is the full width at half maximum of $d<\chi>/dT$. 
It is clear from Eq.(\ref{Omega}) that for a first order transition in
an infinite system $\Omega _c \rightarrow \infty$. Thus, $\Omega _c$
is a quantitative measure to which the folding transition is
cooperative. For finite sized proteins $\Omega _c$ would not be very
large. Analysis of experimental data suggests that $5<\Omega _c<100$
for typical proteins that undergo a two-state transition 
\cite{KlimThirum98coop}.   For
sequence A we find $\Omega _c=5.3$, which given the small size of a model
protein suggests a highly cooperative two-state transition. 

The
characteristic temperatures for sequence B (see Fig. (2b) for the ground
state) are as follows. The collapse
transition takes place at $T_\theta=0.44$, at which $C_v$ has a peak
(or 0.48 if the location of the maximum in $d<R_g>/dT$ is considered). The
folding temperature $T_F$ is found to be 0.30. Thus, sequence B has 
the value of $\sigma =0.32$. It is important to point out that the
cooperativity index $\Omega _c$ for sequence B is significantly smaller  being
equal to 2.0. 

According to the foldability principle the fast folding
sequences, i.e., those which fold thermodynamically and kinetically
by a cooperative two-state transition, 
are characterized by small values of $\sigma $.  In
contrast, sequences, which fold via intermediates and display low
cooperativity of folding, have large $\sigma $. Thus, we expect that
sequence A should be
a fast (i.e., the native state is reached rapidly) 
two-state folder, while B is likely to be a slow
folder with well populated intermediate(s). 
The data given above indicate 
that sequence A for which $\sigma =0.04$ folds  almost three times more 
cooperatively (as measured by $\Omega _c$) than B does. The lower
cooperativity of folding of B is related to the presence of
intermediates. Indeed, the
thermal distribution of $\chi$ values $h(\chi)$ 
at the  folding transition is 
bimodal for A since only two states, folded, {\bf N}, and unfolded, 
{\bf U},   are significantly populated at $T_F$ (Fig. (3a)). In contrast,
$h(\chi)$ for sequence 
B has several dominant peaks, of
which the middle one located at $\chi \sim 0.3$ is attributed to an 
equilibrium intermediate (Fig. (3b)). 

\noindent
{\em Kinetics:} 
The analysis of folding kinetics of sequences A and B further
highlights the dramatic differences in their folding properties. In order
to calculate the  folding time $\tau _F$ we computed the distribution
of first passage times $\tau _{1i}$ from hundreds of individual
folding trajectories. The first passage time $\tau _{1i}$  is the
first time a given trajectory labeled $i$ reaches the native
conformation starting from initially unfolded state. 
If the distribution of $\tau _{1i}$ is known it
is straightforward to calculate the fraction of unfolded molecules
$P_u(t)$ which have not yet reach the native state at time $t$. The
profiles of $P_u(t)$ can be accurately fit with exponentials that
provides a reliable method for computing $\tau _F$ as $\tau_F=\int_0^\infty
P_u(t)dt$. We have also calculated the time dependence of the radius
of gyration $<R_g(t)>$ averaged over multiple folding
trajectories. 

For sequence A we found that the function $P_u(t)$ decays exponentially
over a wide temperature range $T \ge 0.85T_F$. For example, Fig. (4)
shows that  
at $T=0.94T_F$  $\tau _F =2.07\times 10^6$ MCS and folding is
two-state, i.e., $P_u(t) \sim (1-e^{-t/\tau_F})$.  
The time dependence of the 
radius of gyration reveals two collapse stages - the initial
extremely rapid
burst phase followed by  a more gradual chain 
compaction (Fig. (4)). 
Consequently, $<R_g(t)>$ is well fit with
the sum of two exponentials with the times scales  
$\tau_1 = 0.00831\times 10^6$ MCS and $\tau_2 = 0.698 \times 10^6$ MCS. 
We interpret  $\tau_2 $ as the overall time scale for
compaction, i.e., $\tau_2 = \tau_c$, while the first time scale
$\tau_1=\tau_{bc}$ is identified with the burst phase. 
From this, we obtain that 
the ratio of folding and collapse time scales $\tau_F/\tau_c$ is 3.0. 

One of the questions of great current experimental interest
\cite{Eaton99,Roder99} is the
precise relationship between the dynamics of chain collapse (as
measured by the decay of $<R_g(t)>$ with the characteristic time $\tau
_c$) and the time for acquisition of the native conformation, namely,
$\tau _F$. It has been suggested that non-specific collapse of the chain
(this happens for sequences with large $\sigma $) inevitably slows
down 
folding \cite{Thirum95}. Sequence A collapses rapidly and the
subsequent folding is also fast. This seemingly puzzling observation
may be understood in terms of theoretical arguments \cite{Thirum95} and earlier
computations on lattice models without side chains. We had shown that
as long as the initial collapse leads to native-like structure 
\cite{Cam93,Thirum95} (i.e., this is a 
specific collapse), then subsequent folding can be rapid. This is
precisely what takes place in sequence A. Recent experiments have also
come to similar conclusions \cite{Eaton99,Roder99}. 
The ratio $\tau_F/\tau_c$ will depend on
$\sigma $ and external conditions, but for sequences which undergo
specific collapse we expect $\tau_F/\tau_c < 10$.

Sequence B folds on  dramatically  longer time scales.  For example, 
$\tau_F$ at $T=T_F$ is $106\times 10^6$ MCS. 
For comparison, for sequence A $\tau _F$  
at $T_F$ is only $1.84\times 10^6$ MCS. Therefore, sequence A folds 60
times faster than B. This finding is in accord with the values of
$\sigma $, which for A and B
are 0.04 and 0.32, respectively. This exercise illustrates the
validity of the foldability principle based on $\sigma $: The eight-fold
difference in $\sigma $ values between  these sequences results in
{\em sixty-fold} difference in their folding times. Clearly, the description of
folding of these two sequences is given here for illustration purposes
and full validation of 
the foldability principle requires a study of much larger
database of sequences. This has been done in a number of 
papers \cite{KlimThirum96PRL,Veit}, 
in which exhaustive examination of many sequences, both for
lattice and off-lattice   models, offered a reliable statistical proof of the
foldability principle.

The foldability criterion, illustrated using two sequences A and B, 
gives the factor that determines the sequences that are biologically
competent. From the mapping of sequence space to structure space we
found that a large number of sequences can have similar topology for
the native conformation. However, the requirement of stability and
kinetic accessibility of the native conformation further restricts the
number of biologically competent sequences. It is remarkable that a
single sequence dependent thermodynamic parameter $\sigma $ (which is
experimentally measurable) controls the foldability of protein
sequences.

\section{Conclusions}

In this article we have used simple models to show how stable (under
folding conditions) and fast folding protein structures could have
evolved. By merely studying the mapping between sequence space and
structure space (with the focus on minimum energy structures)  
it is easy to understand why only a small fraction of
sequences can have protein-like characteristics. The importance of
using low energy as a natural selection scheme has also been
emphasized by Abkevich {\em et. al} \cite{Shakh96PNAS}. 
As pointed out by these authors it is likely that in pre-biotic
evolutionary process the simple requirement of solubility in water may
have inhibited the synthesis of sequences that are highly
hydrophobic. The arguments developed here and elsewhere
\cite{CamThirum93PRL}  show that indeed the optimal fraction of
hydrophobic residues should not be significantly greater than 0.5. 
Such an optimal value not only ensures the  stability of  folded
structures in water, but also greatly reduces the number of minimum energy
structures.   
For biological viability the dual requirement of
stability and kinetic accessibility is vitally important. The
intrinsic sequence dependent parameter $\sigma $ (see
Eq. (\ref{sigma})) determines foldability of sequences. By this two
step analysis we are able to rationalize how through evolutionary
pressure nature may have solved the Levinthal paradox in sequence
space, i.e., the navigation of sequence space to synthesize
functional proteins. The functional requirement, namely, that certain
residues have precise positions in the folded state of proteins may
further reduce the number of sequences that are biologically
competent. We represent the reduction scheme 
in Fig. (5). 
The number of generations required for a  random sequence to evolve to
functionally competent proteins is an interesting questions. In
principle, this can be addressed using simple caricatures of proteins
\cite{Sasai}. 

\acknowledgments
We are grateful for Oksana Klimov for producing Fig. (5). This work
was supported by a grant from the National Science Foundation.

\newpage
\centerline{FIGURES}
\vspace{0.35cm}

\noindent
{\bf Fig. (1)} Scaling of the number of MES $C$(MES) (squares) is
shown for two
values of the hydrophobic parameter $B_0$, 0 (panel (a)) and -0.1
(panel (b)) and $\Delta =0.6$. Data are obtained 
for the cubic lattice. The pairs of squares for each $N$
represent the quenched averages for different samples of 30
sequences. The number of compact structures $C$(CS) and self-avoiding
conformations $C$(SAW) are also displayed to underscore the dramatic
difference of scaling behavior of  $C$(MES) and $C$(CS) (or
$C$(SAW)). It is clear that for both values of $B_0$ $C$(MES) remains
flat, i.e. it grows no faster than $ln N$.

\noindent
{\bf Fig. (2)} Native structures of sequences A and B are shown in the
panels (a) and (b), respectively. Dark grey side chains indicate
hydrophobic residues, whereas light  grey ones represent hydrophilic
amino acids. Both of the conformations are compact with
well-defined hydrophobic core. 
Figure is generated using program RasMol \cite{RasMol}.

\noindent
{\bf Fig. (3)} Thermally weighted distributions of states $h(\chi )$
quantified using the overlap function $\chi $ at $T=T_F$ for sequences
A (panel (a)) and B (panel (b)). The
profiles $h(\chi )$ are strikingly different. In the upper panel 
only two states, $\chi
\sim 0.6$ (unfolded) and $\chi < 0.2$ (folded), are populated and 
consequently $h(\chi )$ is bimodal. The profile $h(\chi )$ for B
indicates the existence of the third populated intermediate state at
$\chi \sim 0.3$. Accordingly, sequences A and B are classified as two-
and three-state folders, respectively. 

\noindent
{\bf Fig. (4)} The time dependence of the normalized radius of gyration
$<R_g(t)>$ and the fraction of unfolded molecules $P_u(t)$ for
sequence A at $T=0.94T_F$. Data are averaged over 100 (for $<R_g(t)>$)
or 600 (for $P_u(t)$) trajectories. $P_u(t)$ decays exponentially
with the time scale $\tau_F=2.07\times 10^6$ MCS. The approach of
$<R_g(t)>$ to equilibrium is well described by two exponentials with
the times scales $\tau_{bc}=0.083\times 10^6$ MCS and
$\tau_c=0.698\times 10^6$ MCS, where $\tau_{bc}$ corresponds to
extremely rapid burst phase and $\tau_{c}$ manifests more gradual final
compaction. The ration $\tau_F/\tau_{c}$ is approximately 3.0. 
  
\noindent
{\bf Fig. (5)} Schematic illustration of the stages in 
the drastic reduction of
sequence space in the process of evolution to functionally competent
protein structures.

\newpage

\begin{center}

\begin{minipage}{15cm}
\[
\psfig{figure=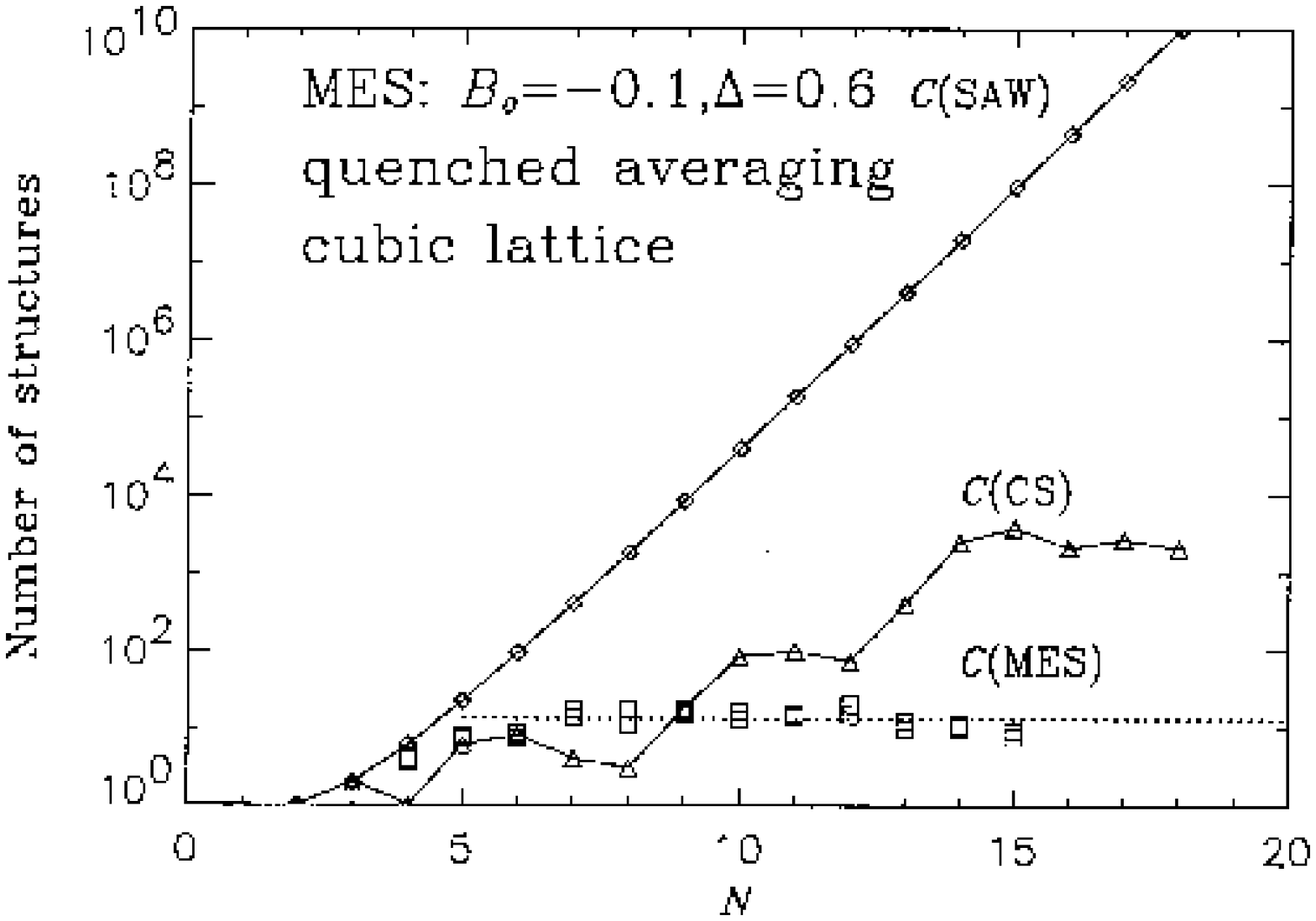,height=9cm,width=12cm}
\]
\end{minipage}

{\bf Fig. 1a} 
\end{center}

\begin{center}

\begin{minipage}{15cm}
\[
\psfig{figure=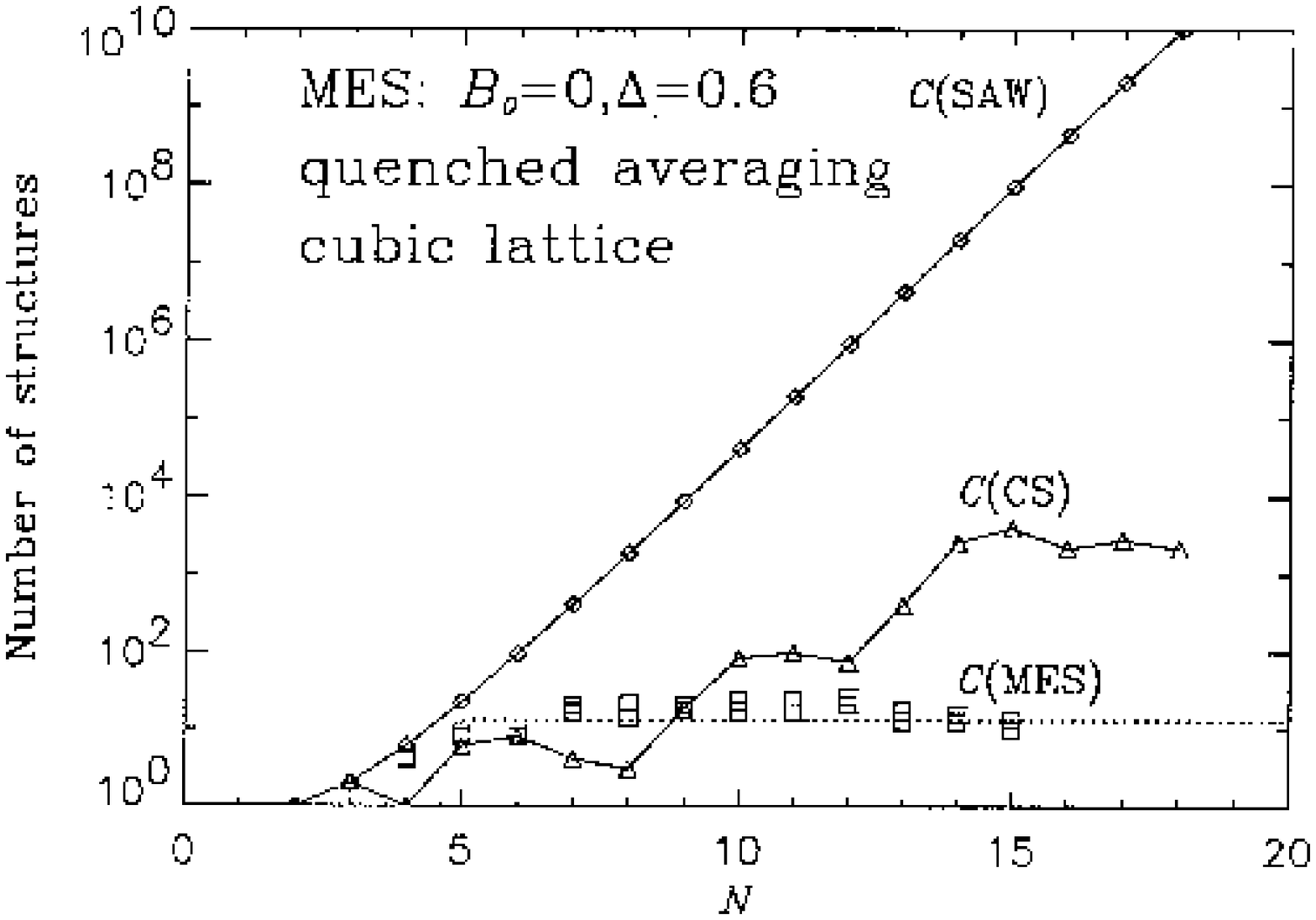,height=9cm,width=12cm}
\]
\end{minipage}

{\bf Fig. 1b} 
\end{center}

\newpage

\begin{center}

\begin{minipage}{15cm}
\[
\psfig{figure=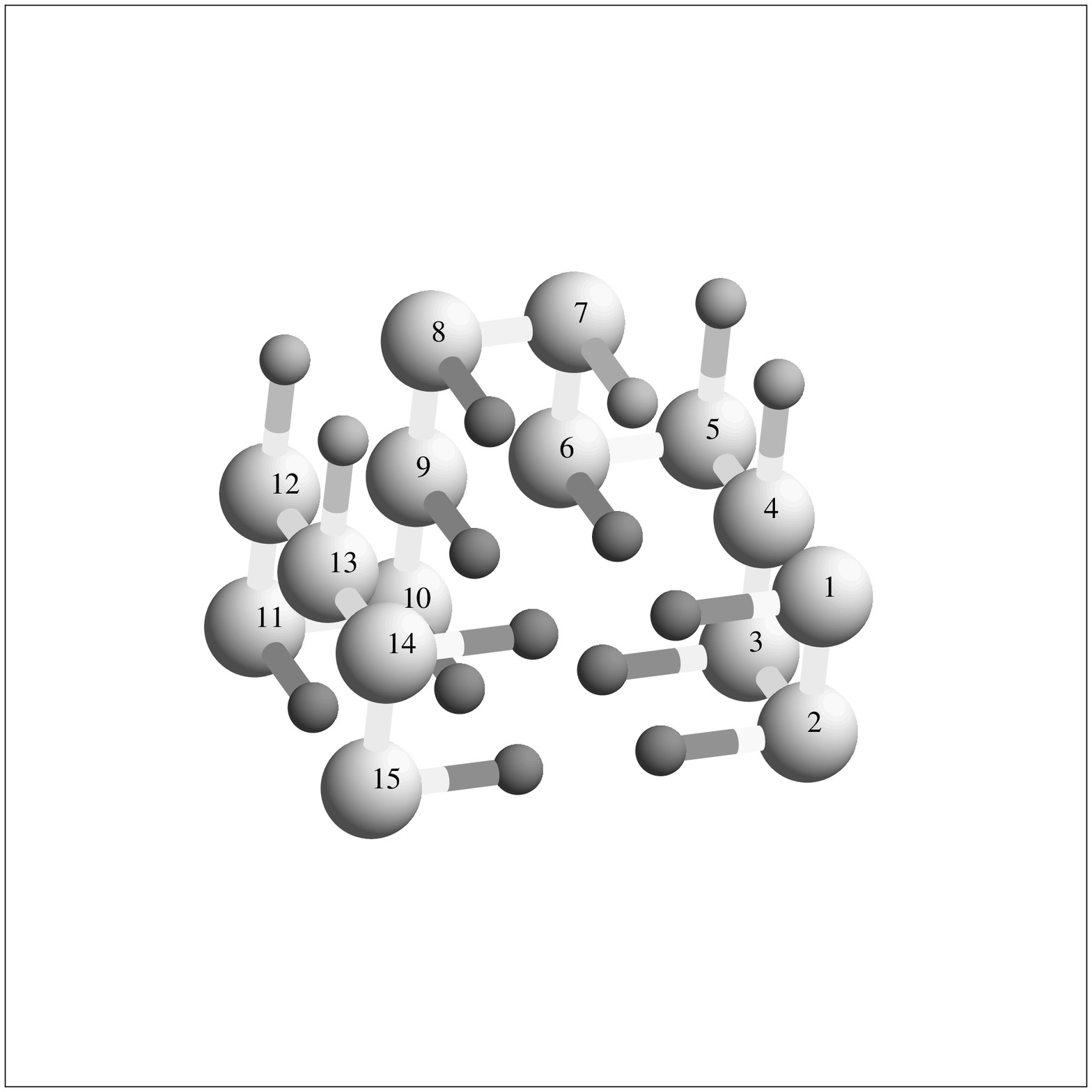,height=9cm,width=9cm}
\]
\end{minipage}

{\bf Fig. 2a} 
\end{center}

\begin{center}

\begin{minipage}{15cm}
\[
\psfig{figure=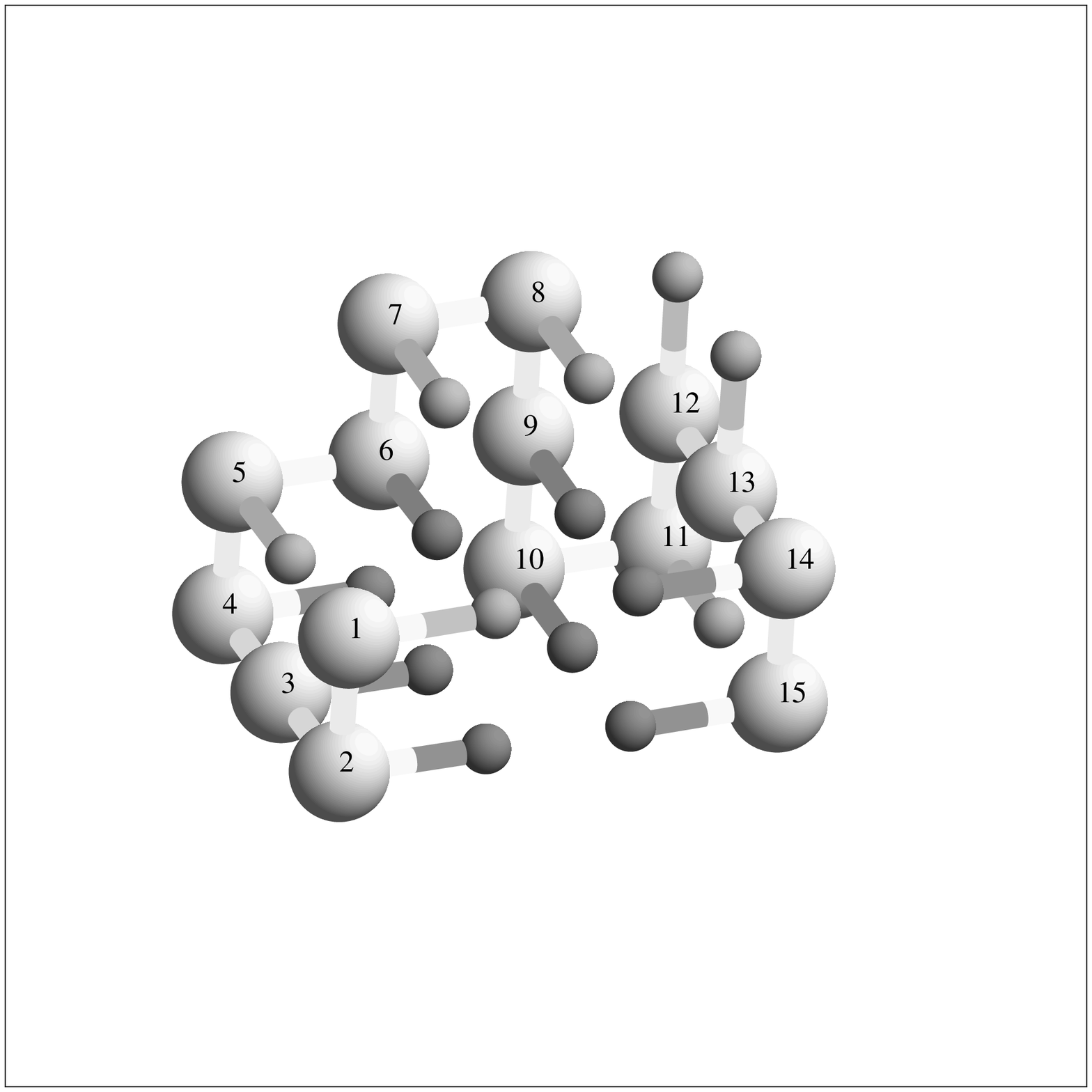,height=9cm,width=9cm}
\]
\end{minipage}

{\bf Fig. 2b} 
\end{center}

\newpage

\begin{center}

\begin{minipage}{15cm}
\[
\psfig{figure=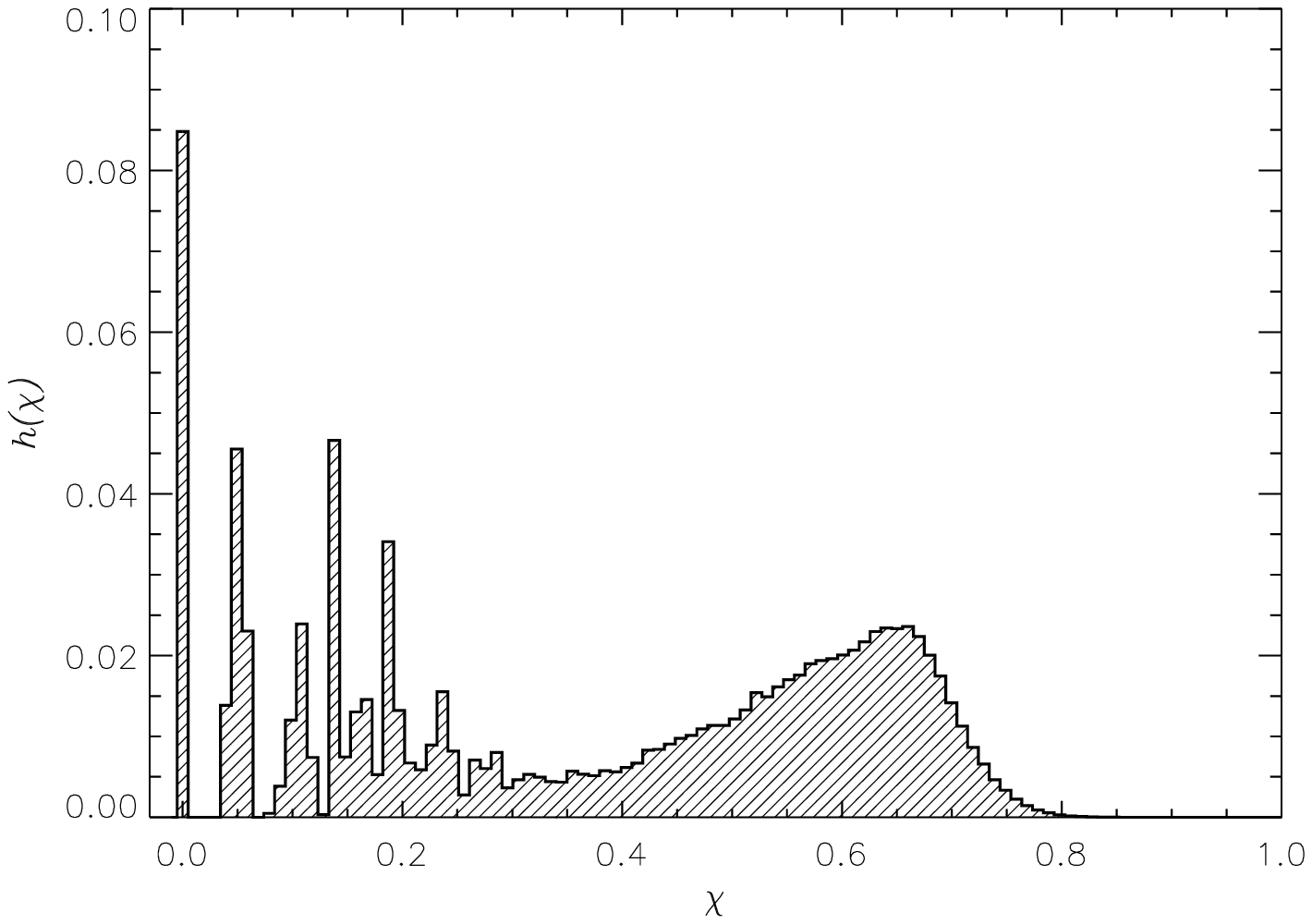,height=9cm,width=12cm}
\]
\end{minipage}

{\bf Fig. 3a} 
\end{center}

\begin{center}

\begin{minipage}{15cm}
\[
\psfig{figure=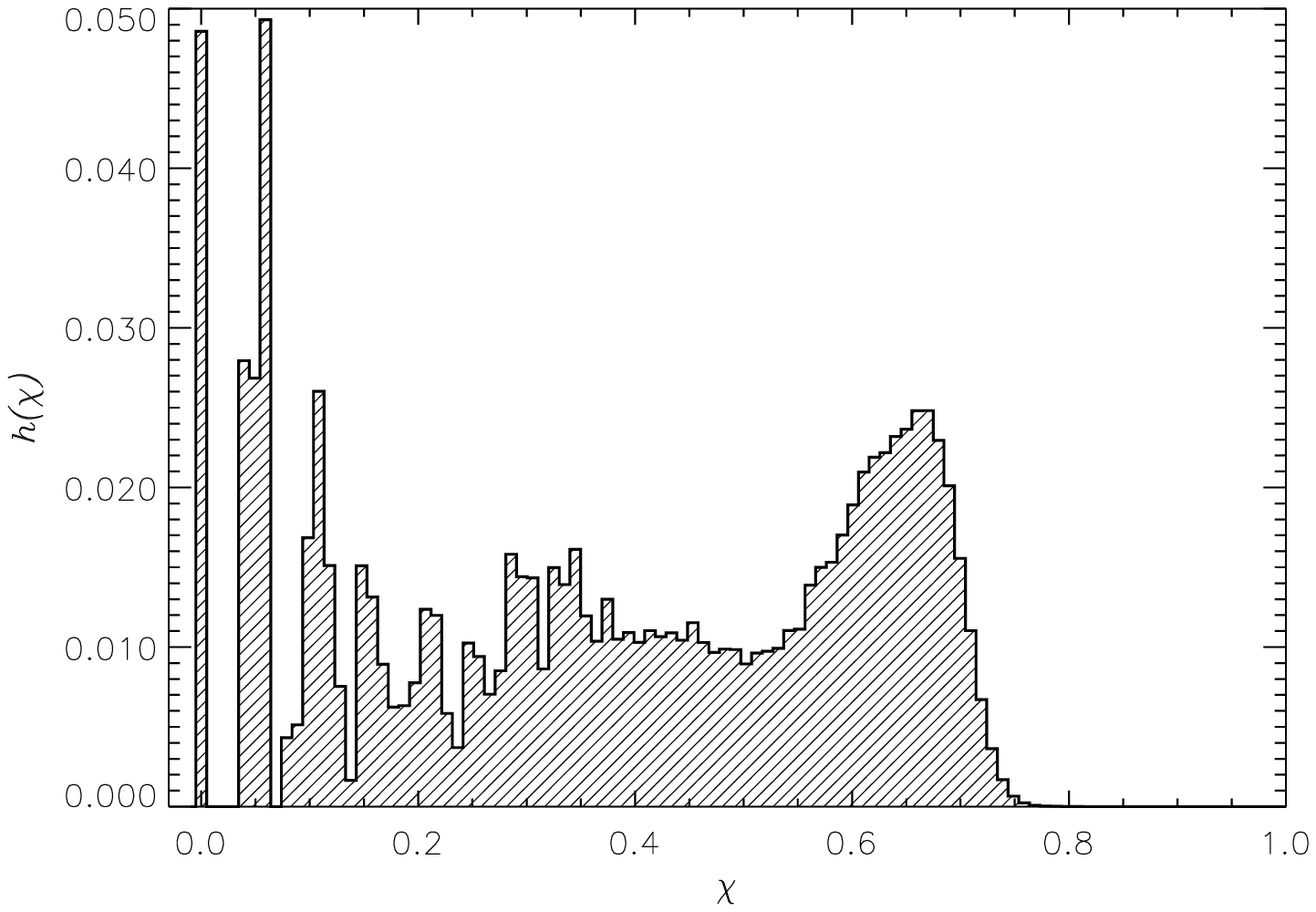,height=9cm,width=12cm}
\]
\end{minipage}

{\bf Fig. 3b} 
\end{center}

\newpage

\begin{center}

\begin{minipage}{15cm}
\[
\psfig{figure=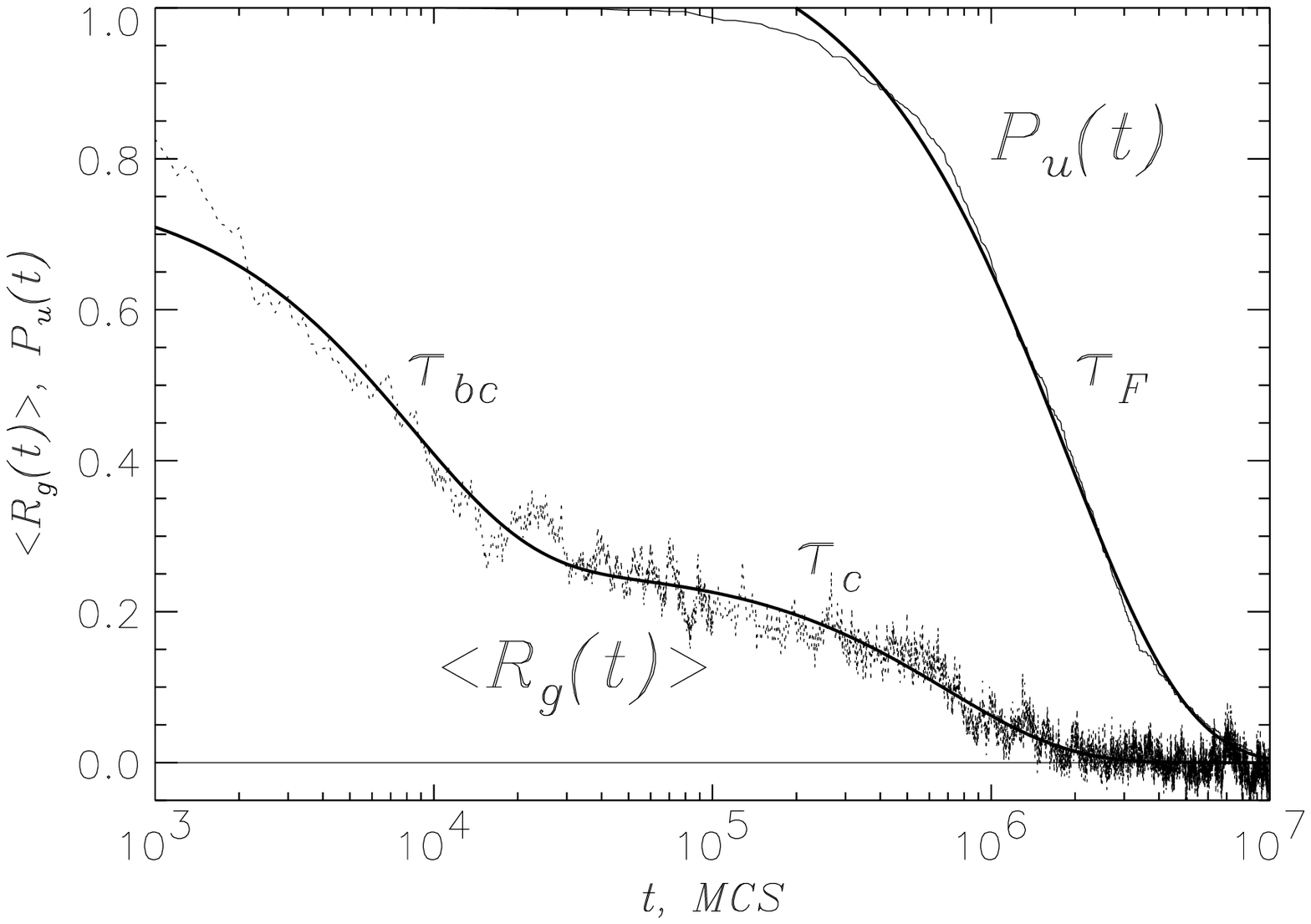,height=9cm,width=12cm}
\]
\end{minipage}

{\bf Fig. 4} 
\end{center}

\newpage

\begin{center}

\begin{minipage}{15cm}
\[
\psfig{figure=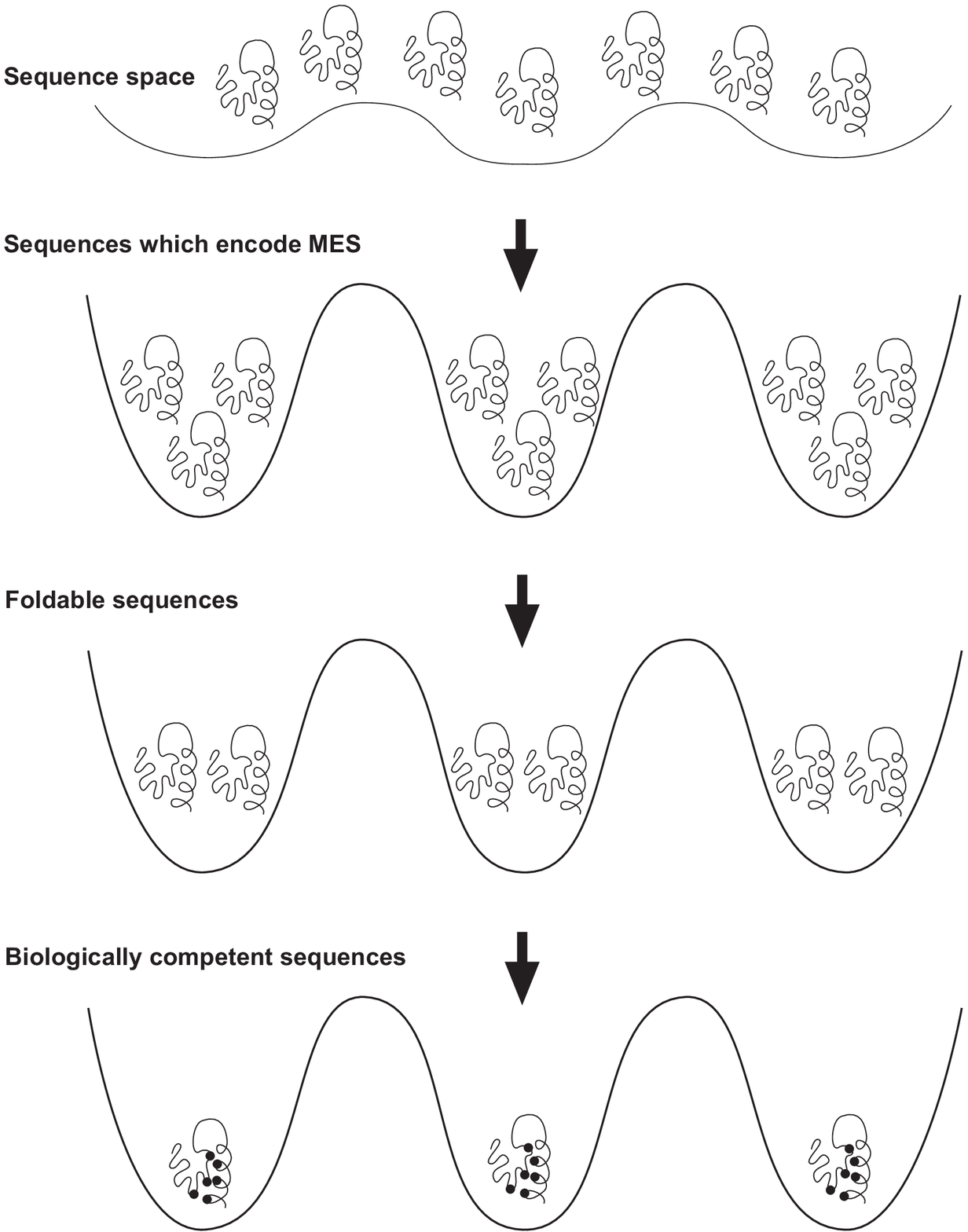,height=20cm,width=16cm}
\]
\end{minipage}

\vspace{0.5in}

{\bf Fig. 5} 
\end{center}

\end{document}